\documentclass[runningheads]{llncs}
\usepackage{graphicx}
\usepackage{amsmath}

\begin{document}

%
\title{Understanding Self-Directed Learning in an Online Laboratory}

\author{Sungeun An\inst{1} \and
Spencer Rugaber\inst{1} 
Jennifer Hammock\inst{2} \and
Ashok K. Goel\inst{1}
}
\authorrunning{S. An et al.}
%
\institute{School of Interactive Computing, Georgia Institute of Technology, Atlanta GA 30308, USA \and National Museum of Natural History, Smithsonian Institution, Washington, D.C. 20002, USA
\\
\email{sungeun.an@gatech.edu}}

%
%
%
%
\maketitle              
\begin{abstract}

Many studies on using online laboratories for learning focus on pedagogical contexts in K-12 education with well-defined problems as well as well-defined learning goals, assessments, and outcomes. We describe a study on the use of an online laboratory for self-directed learning through the construction and simulation of conceptual models of ecological systems. The learning goals and the demographics of the learners in this study are unknown; only the modeling behaviors and outcomes are observable. We analyzed the modeling behaviors of 315 learners and 822 instances of learner-generated models using machine learning techniques such as clustering and dimensionality reduction. We found three types of learner behaviors: observation (focused on simulation), construction (focused on conceptual model), and exploration (full cycle of model construction, simulation, and revision). We found that while the observation behavior was most common, the exploration behavior led to higher model quality.

\end{abstract}
\section{Introduction}
In recent years, self-directed learning through media and information technology has become increasingly prevalent. For example, people often learn by reading and contributing to articles in Wikipedia \cite{forte2006wikipedia} and by using and producing open-source software in Scratch \cite{scaffidi2012skill}. This is learning without any instructor, syllabus, or mandate to cover essential materials. It is what Haythornthwaite et al. (2018) call ``Learning in the Wild'' (with due acknowledgement of Hutchins' original ``Cognition in the Wild'') \cite{haythornthwaite2018learning}. It is non-formal learning taking place outside classroom settings by asking, answering, and learning at the discretion of the learners.

Online laboratories too provide affordances for self-directed learning. A learner can learn independently and investigate their own hypothesis by designing multiple experiments and evaluating their hypothesis via simulations at the discretion of the learner \cite{van2005co}\cite{joyner2014mila}. Learners can use online laboratories to represent various phenomena around the learners and better understand the phenomenon or predict the outcomes of hypothetical changes \cite{van2005co}\cite{joyner2014mila}. Given that addressing environmental problems is among the biggest challenges, learners can play a more active role in learning about ecological phenomenon using online laboratories.

One challenge in using online laboratories for self-directed learning is measurement of the learning outcomes as there almost surely will be a large variance in the ecological phenomena being modeled as well as in the learning goals and behaviors. However, at present there is a lack of understanding of the processes and outcomes of self-directed learning in online laboratories including in the domain of ecology. Many studies on the use of virtual laboratories for learning focus on pedagogical contexts in K-12 education with well-defined problems as well as well-defined learning goals, assessments, and outcomes (for example, \cite{van2005co}\cite{basu2013ctsim}\cite{gobert2013log}\cite{joyner2014mila}). As online laboratories become increasingly widespread, it is important to not only formulate appropriate measures of learning but also to validate learning theories and findings from the literature.

This paper analyzes self-directed learning in VERA (Virtual Experimentation Research Assistant), a publicly available online laboratory for modeling ecological systems \cite{an2018vera}\cite{an2020scientific} (vera.cc.gatech.edu). VERA is a web application that enables users to construct conceptual models of ecological systems and run interactive agent-based simulations of these models. This allows users to explore multiple hypotheses about ecological phenomena and perform ``what if'' experiments to either explain an ecological phenomenon or attempt to predict the outcomes of future changes to an ecological system.

We investigate three research questions. (1) What kinds of learning behaviors emerge in self-directed learning in an online laboratory? (2) How do the behaviors evolve over time? (3) How do the learning behaviors relate to learners with different engagement levels and model quality? In this study the learning goals, the demographics of the learners or even their precise geographical location are unknown. The only observables are the modeling behaviors and outcomes.

\section{Related Work}

The topic of measuring learning in online laboratories has already found significant traction within the AI in Education community. For instance, Gobert et al. (2013) assessed middle school students’ skills in designing controlled experiments from online microworlds in the Inq-ITS system (Inquiry Intelligent Tutoring System) \cite{gobert2013log}. The authors used students’ log data to hand-score their actions based on pre-defined categories. Basu et al. (2013) assessed middle school students’ learning process and outcomes based on metrics that specify the distances between the student’s models and the corresponding expert models \cite{basu2013ctsim}. These studies are interesting in that they use the actions students take within online laboratories as the basis for measurement \cite{gobert2013log}\cite{basu2013ctsim}\cite{joyner2014mila}. However, the primary setting for these studies was not on self-directed learning, but on well-defined problems in K-12 science education with formal learning and specific goals.

Other studies have extracted learner behaviors for measuring and understanding informal learning, self-regulated learning, and self-directed learning in open exploratory environments \cite{blikstein2014programming}\cite{hu2017learning}\cite{maldonado2018mining}. Some researchers have analyzed observed behaviors to uncover learning patterns in MOOCs \cite{maldonado2018mining} and an online problem-solving environment \cite{hu2017learning}, typically with clustering methods. However, these approaches were used in contexts where learning outcomes could be measured (e.g., video completion, solutions to given problems). 


Perhaps the closest literature to our research is recent work on clustering learner data \cite{desmarais2013clustering}\cite{hu2017learning}\cite{maldonado2018mining} and assessing informal learning using Scratch \cite{yang2015uncovering}\cite{scaffidi2012skill}\cite{matias2016skill}. On the one hand,  some of our learning techniques build on this earlier work. On the other, the two context are very different: informal learning about beginner programming by children in earlier work, and self-directed learning in an online laboratory for ecological modeling in the present study.

\section{A Brief Description of VERA}

In VERA, learners build conceptual models of complex phenomena, evaluate them through agent-based simulation, and revise the models as needed. VERA uses Component-Mechanism-Phenomenon (CMP) language \cite{joyner2011evolution}, a variation of the Structure-Behavior-Function (SBF) language \cite{goel2009structure}, for conceptual modeling of complex systems. VERA provides the syntax and semantics to construct the CMP conceptual models of an ecological phenomenon consisting of interacting components (e.g., biotic, abiotic), relationships (e.g., consume, destroy, etc.), and their properties (e.g., initial population, lifespan, etc.). Following our earlier work on the ACT and MILA-S system \cite{vattam2011behavior}\cite{joyner2014mila}, VERA uses an artificial intelligence compiler to automatically translate the patterns in the conceptual models into the primitives of agent-based simulation of NetLogo \cite{wilensky2006thinking}.  

\begin{figure}[ht!]
\centering
\includegraphics[width=0.8\textwidth]{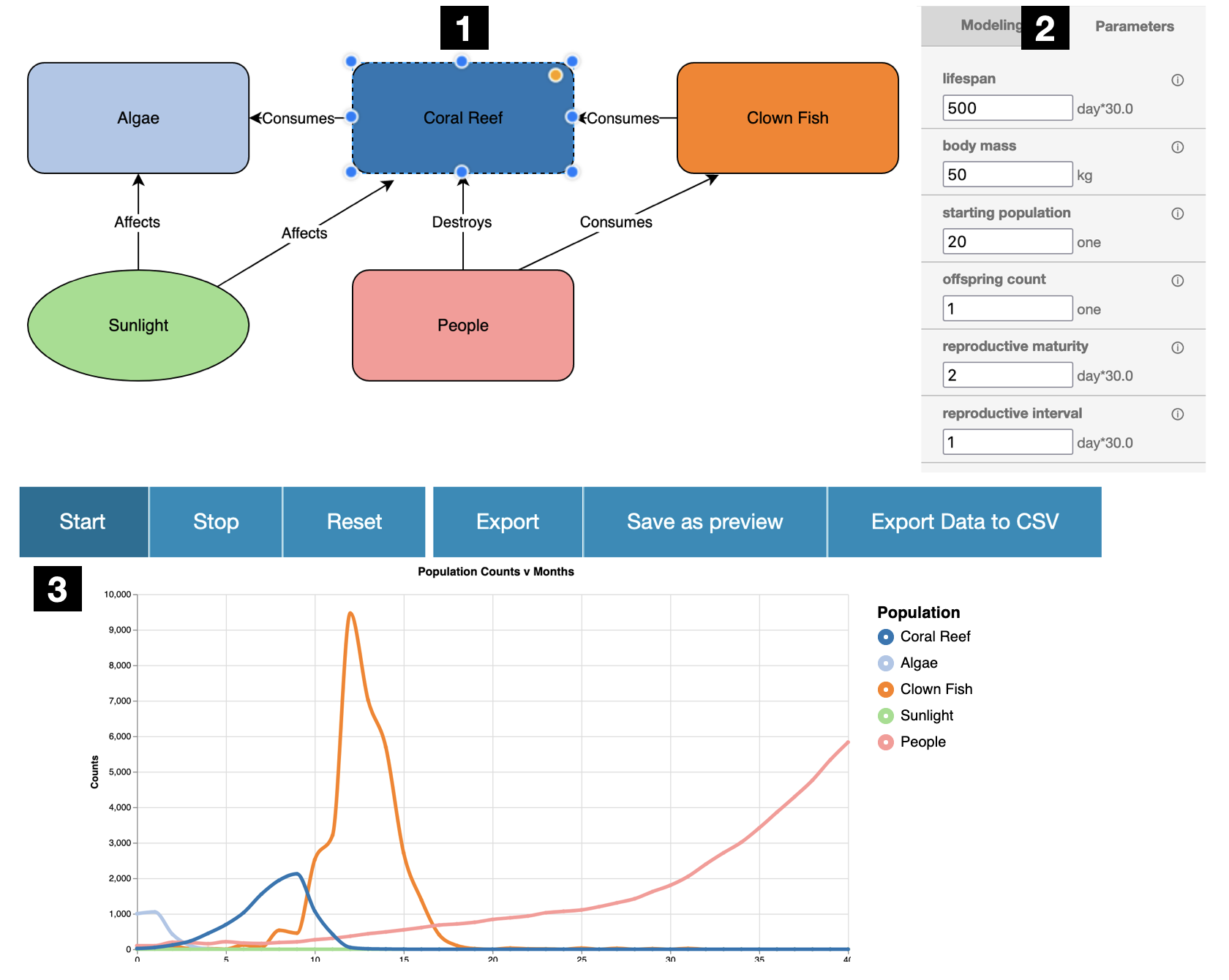}
\caption{An Example of Conceptual Model Created by a Self-Directed Learner. (1) Model Construction. (2) Parameterization. (3) Simulation. }
\label{fig:vera_model}
\end{figure}

Figure \ref{fig:vera_model} shows an example of a conceptual model created by a self-directed learner (1), and the corresponding simulation output (3). Learners can add biotic/abiotic components and relationships between them in a canvas, change a set of simulation parameters for each component/relationship, and run agent-based simulation of the conceptual model. Learners' log data within the VERA system creates timestamped records of actions such as adding a component, removing a component, or connecting two components with a relationship. These individual actions were categorized into three activity classes: model construction, parameterization, and simulation.

\begin{enumerate}
    \item \textit{Model Construction}: is defined as instances wherein something new is inserted into the model (both components and relationships), or a previously existing portion of the model is removed.  
    \item \textit{Parameterization}: is defined as instances where previously-added components or relationships’ parameters are modified.
    \item \textit{Simulation}: is defined as instances wherein simulation is executed.
\end{enumerate}

\section{Analysis of Self-Directed Learning in VERA}
VERA has been publicly available and globally accessible through multiple websites since the fall of 2018. A preliminary analysis showed that several thousand users accessed VERA through the end of 2021. However, a large majority of users spent only very limited time with VERA and did not construct or simulate any model. Only 315 learners during this time constructed and/or simulated a total of 822 models. We deliberately did not collect any data on the demographics or the precise geographical location of the learners to protect their privacy. Instead, only two types of data are available for analysis: \textit{(1) Learners’ interactions with VERA (log data)}; and \textit{(2) Learners’ final work products (conceptual models).}

\subsection{Behavioral Patterns}
We analyzed learners’ modeling behaviors in three phases. First, an activity sequence for each model was created. Second, each activity sequence was grouped by their sequence lengths (short, medium, long). Third, a Levenshtein Distance algorithm was applied to each group to measure similarities among the sequences \cite{levenshtein1966binary} and then the sequences were clustered using the Agglomerative Hierarchical method \cite{jain1988algorithms}.


\subsubsection{Creating Activity Sequence}
In order to identify the behavior patterns, we first extracted action sequences for every model of a learner \textit{l}. For instance, if a learner performed a series of actions of adding a component, adding another component, and running a simulation within his or her model, the activity sequence is ‘ccs’ (construction, construction, simulation). Given that an activity has no time duration in our data, we focus on the transition from one activity to another. Having the segments of time may result in some activities displayed if the granularity of the time segment is longer than the time between activities. This leaves a total of 822 activity sequences, one for each model, created by the 315 learners. Mean and median sequence length are respectively 52.17 and 15 with a minimum length of 1 and a maximum length of 1605.


\subsubsection{Segmenting Activity Sequence}
Levenshtein Distance computes the number of insertions, deletions and replacements needed to transform one string into the other \cite{levenshtein1966binary}. This means the similarity for two sequences with the same pattern can be measured as low when they have significantly different sequence lengths. For this reason, we grouped similar sequence lengths together. First, outliers (too short or too long sequences) that are above a threshold of mean + 2*SD and below the threshold of mean - 2*SD were eliminated (N=33). Then a segmentation optimization method (Kernel Density Estimation (KDE)) was applied to split the remaining 789 sequences into three different groups by their length based on two local minima in density: short sequence ($<$29.59, N=556), medium sequence ($\geq$29.59, $<$42.85, N=76), and long sequence ($>$42.85, N=157).

\begin{figure}[ht!]
\centering
\includegraphics[width=\textwidth]{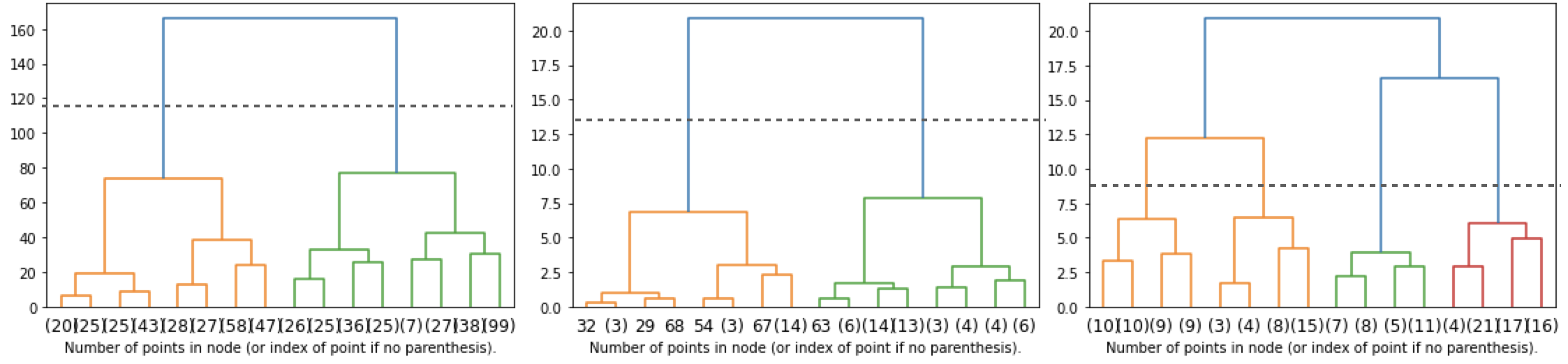}
\caption{Dendrograms with data points (x-axis) and cluster distance (y-axis) of (1) Short Group (Left) (2) Medium Group (Middle) (3) Long Group (Right).}
\label{fig:dendrogram}
\end{figure}

\subsubsection{Behavior Clustering of Similar Sequences}
The extracted activity sequences from the data are divided into three length groups, and the Levenshtein Distance (a string metric for measuring the difference between two sequences) was applied within each length group \cite{levenshtein1966binary}. An Agglomerative Hierarchical method, the most common type of hierarchical clustering to group objects in clusters based on their similarity, is used to aggregate the most similar sequences based on the Levenshtein distance matrix \cite{jain1988algorithms}\cite{desmarais2013clustering}. The bottom-up algorithm initially treats each sequence as individual cluster and then successively merge pairs of clusters. 

\begin{figure}[ht!]
\centering
\includegraphics[width=0.9\textwidth]{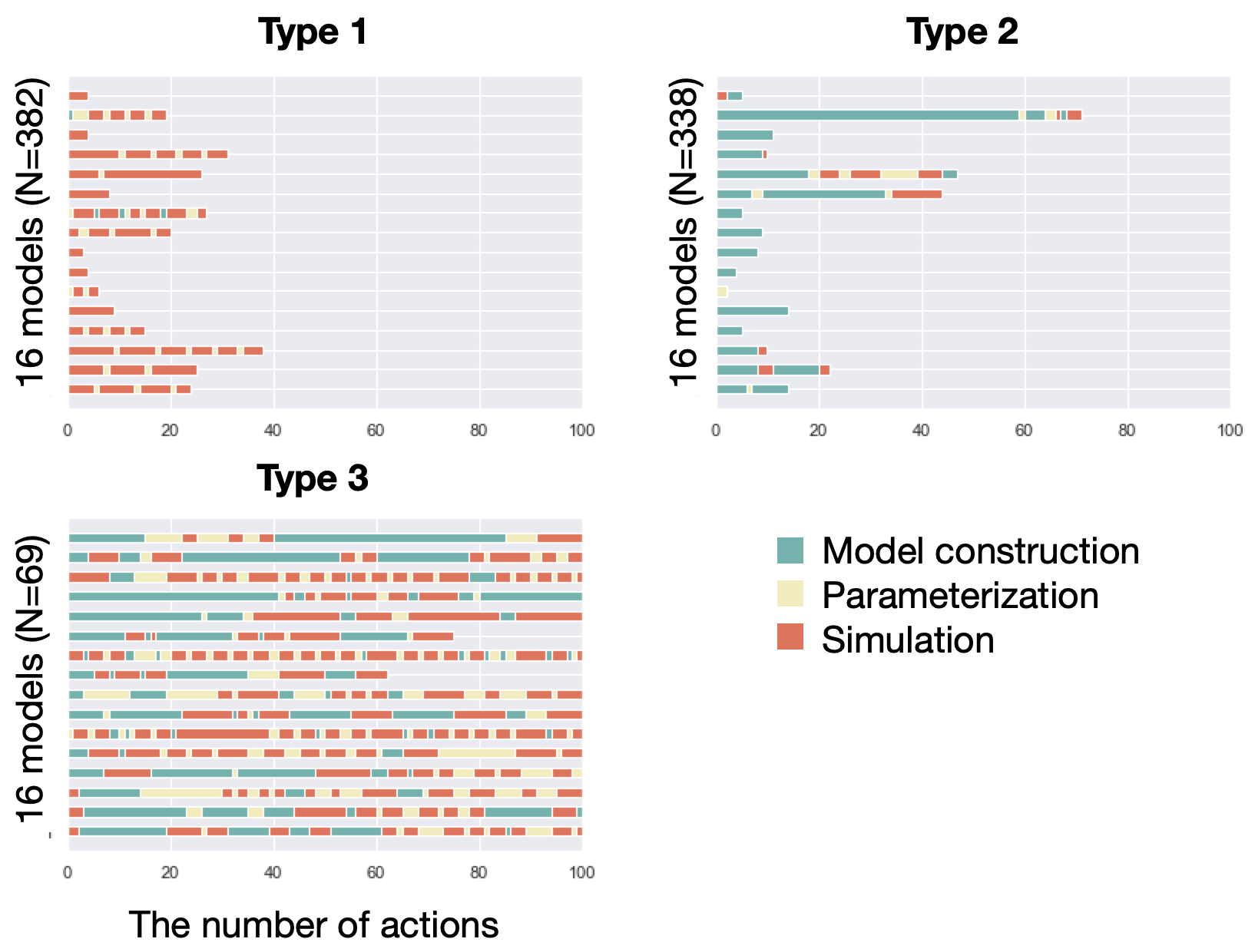}
\caption{Three Behavior Clusters of Similar Activity Sequences. 16 activity sequences are randomly selected for each type. Type 1 focuses on experimenting with very little or no model construction. Type 2 focuses on model construction. Type 3 shows various activities.}
\label{fig:model_cluster}
\end{figure}

As a result, the short length group produced two sequence clusters, the medium length group produced two sequence clusters, and the long length group produced three sequence clusters, as illustrated by the Dendograms in Figure \ref{fig:dendrogram}. A total of seven clusters were visually compared and merged into three clusters; Figure \ref{fig:model_cluster} illustrates the resulting three clusters using the visualization technique in \cite{desmarais2013clustering}.

\subsubsection{Behavioral Clusters}
Figure \ref{fig:model_cluster} shows three behavioral clusters in VERA with 16 randomly selected example sequences for each cluster. Each horizontal line in the figure shows a sequence of activities in a model, the length of an activity in a sequences corresponds to the frequency of the activity. The sequence clusters have the following characteristics:  


\begin{enumerate}
    \item \textbf{Type 1} (N=382): \textit{Observation.} The learners engage in experimenting with different simulation parameters with very little or no evidence of construction of conceptual models.
    \item \textbf{Type 2} (N=338): \textit{Construction.} The learners engage in short sessions of model construction with little or no simulation of the conceptual models.
    \item \textbf{Type 3} (N=69): \textit{Exploration (or Full Cycle).} The learners engage in a full cycle of model construction, parameterization, and simulation.
\end{enumerate}

\subsection{Learners' Engagement}
\subsubsection{Aggregated Learner Data}\label{learner_data}
We create a set of five features using the aggregated user data to identify different learner types: \textit{1) the number of original models, 2) the number of copied models, 3) total counts of model construction, 4) total counts of parameterization, and 5) total counts of simulations.} For each learner \textit{l}, we aligned all models of a learner \textit{l} including original and copied models in sequence from the earliest to the latest. Then, we constructed a \textit{i} x 3 matrix \begin{math}P_{l}\end{math} using the learners' models and frequencies of their three activities (model construction, parameterization, simulation), where \begin{math} f_{i, j} \end{math} is the frequency of activity \begin{math} a_{j} (1 <= j <= 3) \end{math} in model \begin{math} i.\end{math}

\begin{equation} \label{eq1}
P_{l} = 
\begin{pmatrix}
f_{1,1} & f_{1, 2} & f_{1, 3} \\
\vdots & \vdots & \vdots \\
f_{i,1} & f_{i, 2} & f_{i, 3}
\end{pmatrix}
\end{equation}

\noindent We compute \begin{math}P_{l}\end{math} for all 315 learners to construct a matrix \begin{math}T_{u}\end{math} that has 315 x 5 dimensions. 

\begin{equation} \label{eq2}
T_{l} = 
\begin{pmatrix}
v_{1, 1} & v_{1, 2} & v_{1, 3} & v_{1, 4} & v_{1, 5} \\
\vdots & \vdots & \vdots \\
v_{l, 1} & v_{l, 2} & v_{l, 3} & v_{l, 4} & v_{l, 5} \\
\end{pmatrix}
\end{equation}

\noindent where \begin{math}v_{l, 1}\end{math} is the number of original models a learner \begin{math}l\end{math} made, and  \begin{math}v_{l, 2}\end{math} is the number of copied models for a learner \begin{math}l\end{math}. \begin{math}v_{l, 3}, v_{l, 4}, v_{l, 6}\end{math} are created by averaging columns in \begin{math}P_{l}\end{math}.

\subsubsection{Learner Clustering with PCA and K-Means++}
We used Principal Component Analysis (PCA) as a dimension reduction technique to find a linear combination of the features in \begin{math}T_{l}\end{math} \cite{hotelling1933analysis}. The feature values were scaled as few values with different quantities may impact the linear regression algorithm. 

All learners were plotted with the first and second PCA components (PC1 and PC2). PC1 explains 77.48\% of the variance and PC2 explains 9.53\% of the variance. Together, they explain 87.01\%. Consequently, a new matrix \begin{math}T_{p}\end{math} was created that has 315 x 2 dimensions with the two vectors \begin{math}a_{1}, a_{2} \end{math} that define the first two principal components. 

We applied a clustering algorithm on the \begin{math} T_{p} \end{math}. We used the K-means++ algorithm, which is the regular K-means algorithm with a smarter initialization of the centroid and improves the quality of the clustering. To find an appropriate k, we plotted the total Within-Group Sum of Squared Error (SSE) for increasing k values (see Figure \ref{fig:learner_cluster} (Left)). We chose k=5 as increasing the value of k from 4 to 5 reduces considerably the value of the total Within-Group SSE. However, further increasing k affects the SSE values only minimally. 

\subsubsection{Engagement Learner Groups}
\begin{figure}[ht!]
\centering
\includegraphics[width=0.95\textwidth]{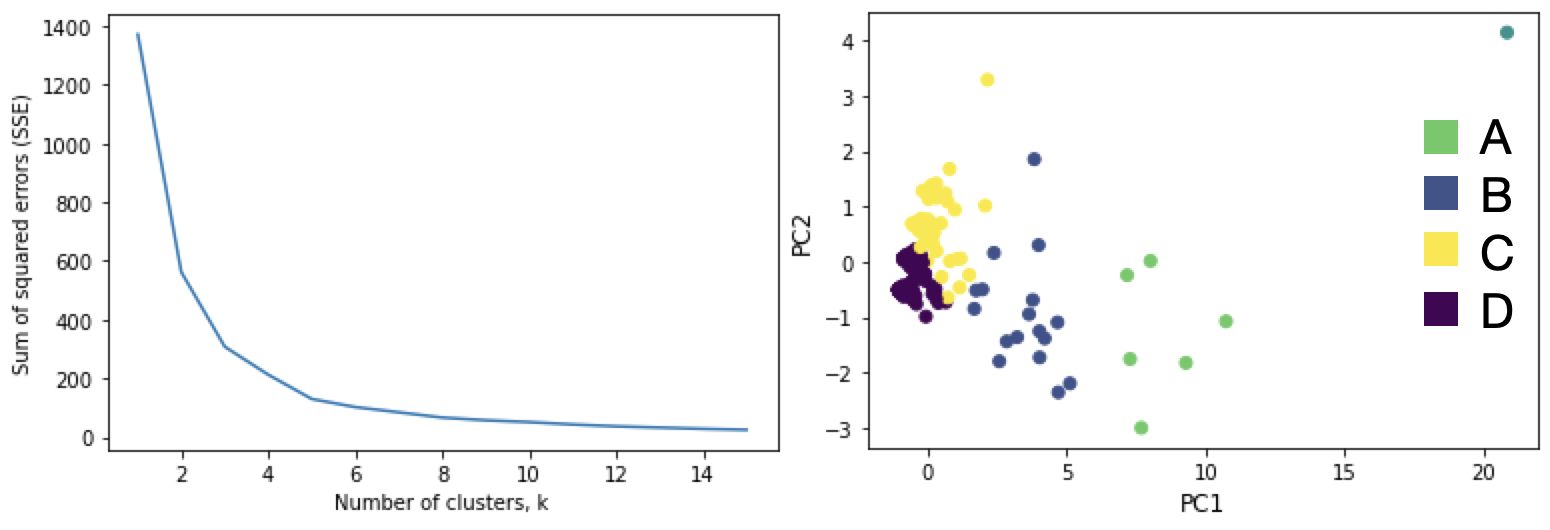}
\caption{The Total Within-Group SSE Values by Different Number of Clusters (K) (Left). K=5
was Selected. The Corresponding Learner Clusters (Right). The rightmost cluster with only one were excluded from the analysis.} \label{fig:learner_cluster}
\end{figure}

Table \ref{tab1} summarizes feature importance by the magnitude of the corresponding values in the eigenvectors. All features in PC1 have similar importance while Feature 3, 4, 5 (\textit{total counts of construction, parameterization, and simulation}) are the most important for PC1. Feature 1 (\textit{the number of original models}) is the most important for PC2. In other words, higher PC1 means a higher level of engagement (many actions). Learners with higher PC1 have completed many actions. Higher PC2 means more models (especially original models), but fewer actions on them. The rightmost cluster with only one were excluded from the analysis (see Figure \ref{fig:learner_cluster}). Therefore, we derived four types of learners as illustrated in Figure \ref{fig:learner_cluster} (Right): 
\begin{enumerate}
    \item Learner Cluster A (green): This learner group has relatively high PC1 and low PC2 values. They are the most active learners who created some conceptual models with many actions on them.
    \item Learner Cluster B (navy blue): This learner group has relatively medium PC1 and medium PC2 values. They are moderately active learners who created some models with moderate actions on them.
    \item Learner Cluster C (yellow): This learner group has relatively low PC1 and high PC2 values. They are less active learners who created many models but with fewer actions on them.
    \item Learner Cluster D (purple): This learner group has relatively low PC1 and low PC2 values. They are less active learners who created less models.
\end{enumerate}

\begin{table}[ht!]
\caption{The Importance of Each Feature \begin{math} v\end{math} by the Magnitude of the Corresponding Values in the Eigenvectors (Higher Magnitude — Higher Importance).}\label{tab1}
\begin{tabular}{|l|l|l|l|l|l|}
\hline
Component &  \begin{math} v_{1}\end{math} & \begin{math} v_{2}\end{math} & \begin{math} v_{3}\end{math} & \begin{math} v_{4}\end{math} & \begin{math} v_{5}\end{math}\\
\hline
PC1 &  0.39588313 &  0.40277977  &    0.46261331  &    0.48078568  &0.48567383 \\
PC2 & 0.83169548 &  0.21416105 & -0.20546292 &  -0.34241763&  -0.32086262 \\ 
\hline
\end{tabular}
\end{table}

\subsection{Behavior Patterns with Different Engagement Learner Groups}
Table \ref{tab1:behavior_learner} summarizes the distribution of behavior types for each learner group. Type 1 (Observation) behavior is the most common in most learner groups except for C. While Type 1 behavior is widely used in all groups, Type 2 (Construction) and Type 3 (Exploration or Full Cycle) sharply divide between the more engaged group (A, B) and the less engaged group (C, D). Type 2 is more commonly used for the less engaged groups whereas Type 3 is more commonly used for more engaged groups. There was a statistically meaningful relationship between behavior types and learner groups as determined by chi-square test (p$<$0.001). 

\begin{table}
\caption{The Distribution of Behavior Types for Each Learner Group. A (Higher Engagement) to D (Lower Engagement). Highest Value in \textbf{Bold}. Percentage in Parenthesis.}\label{tab1:behavior_learner}
\begin{tabular}{|l|l|l|l|}
\hline
Learner Group &  Type 1 (Observation) & Type 2 (Construction) & Type 3 (Full Cycle)\\
\hline
Group A &  \textbf{38} (48.71\%) &  22 (28.20\%) &    18 (23.07\%) \\
Group B &  \textbf{38} (50.00\%)&  19 (25.00\%) &    19 (25.00\%) \\
Group C &  100 (37.73\%)&  \textbf{147} (55.47\%)  &    18 (6.79\%) \\
Group D &  \textbf{168} (52.17\%) &  142 (44.09\%) &    12 (3.72\%)) \\
\hline
\end{tabular}
\end{table}

\subsection{Behavior Patterns with Different Model Quality}

We used two proxies to measure model quality. \textit{Model complexity} is defined as the total number of model components and relationships (referred as \textit{depth} in other literature \cite{matias2016skill}]\cite{scaffidi2012skill}). \textit{Model variety} is defined as the number of unique categories used for components and relationships (commonly referred as \textit{breadth} in other literature \cite{matias2016skill}). 

For model complexity and variety, there is a statistically significant difference between the types as determined by one-way ANOVA test (model complexity: p$<$ 0.001, f$=$ 75.36; model variety: p$<$ 0.001, f$=$26.80). We conducted t-tests for pairwise comparisons between 1) behavior types and model complexity and 2) behavior types and model variety. For behavior types and model complexity, significant differences were found between all pairwise comparisons (e.g., Type 1 and Type 2: p$<$0.005, t$=$2.9835, Type 1 and 3: p$<$0.001, t$=$-7.6527, Type 2 and 3: p$<$0.001, t$=$-11.2651). For behavior types and model variety, significant differences were found between all pairwise comparisons (e.g., Type 1 and Type 2: p$<$0.01, t$=$2.6965, Type 1 and 3: p$<$0.001, t$=$-5.8629, Type 2 and 3: p$<$0.001, t$=$-6.5342).

\begin{figure}[ht!]
\centering
\includegraphics[width=0.95\textwidth]{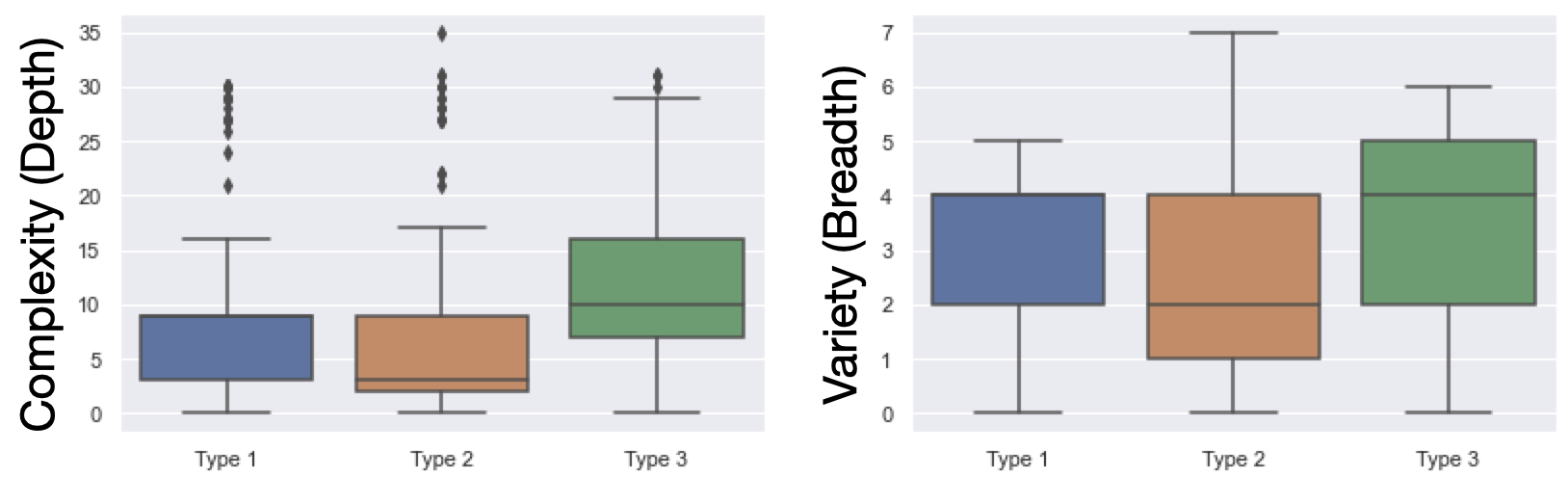}
\caption{Mean, Median, and Lower and Upper Quartiles of Model Complexity (Left) and Variety (Right).} \label{fig:model_quality}
\end{figure}

Figure \ref{fig:model_quality} shows the differences in model complexity and model variety for each behavior type. The conceptual models that manifested Type 3 (full cycle) behavior had the most complex models (mean$=$12.5) followed by Type 1 (mean$=$8.52) and Type 2 (mean$=$6.22). The conceptual models that manifested Type 3 (full cycle) behavior had the most variety models (mean$=$3.5) followed by Type 1 (mean$=$2.9) and Type 2 (mean$=$2.3). 

\section{Discussion of Self-Directed Learning in VERA}


We found three types of self-directed learning behaviors in VERA. Type 1 (Observation) shows a mostly passive behavior in that learners first either simply copy exemplar models or make small tweaks to them, and then run simulations on them. Thus, there is a frequent parameterization-simulation sequence in the data (as illustrated in Figure \ref{fig:model_cluster}). This behavior pattern typically has short session lengths (mean$=$22.62 actions). It is the most common behavior among three learner groups (A,B,D). As Sins (2006) pointed out \cite{sins2005difficult}, modeling is a cognitively difficult process,
and thus observing the simulation of the exemplar models and modifying them can help learners better understand the system being modeled. Previous studies also indicated the importance of learning from exemplar or expert models \cite{basu2013ctsim}. For most (but not all) learners, Type 1 preceded Type 3, and many learners created an original models once they had attained a decent working knowledge of the observed behavior.


Type 2 (Construction) learners spent most time of adding/deleting elements in a model with much less attention to model simulation/evaluation. This behavior pattern has the shortest session length among the three behaviors (mean$=$16 actions) and also less common for more engaged learners (A, B) than less engaged learners (C, D) (see Table \ref{tab1:behavior_learner}. Detailed analysis suggests that this type of behavior typically is found in construction of initial models. For example, 190 out of 286 learners showed this behavior for their first model (followed by 77 (Type 1) and 19 (Type 3)). This aligns well with earlier findings described in the literature on the use of virtual laboratories in pedagogical contexts in K-12 education \cite{joyner2014mila}. The authors found that the initial phase of model construction consisted primarily of testing out the connections between conceptual model and simulations than explicitly trying to accurately model the system \cite{joyner2014mila}. Multiple teams in their study constructed conceptual models and made small changes to them early on to understand the ways in which elements in the conceptual model would manifest in the model simulation.

Type 3 (Exploration) learners display the full cycle of  exploratory behavior consisting of model construction, parameterization, and simulation, and comes closest to the scientific way of thinking described in the literature (for example, by \cite{schwarz2009developing}). This behavior had the longest session among the three behaviors (mean$=$154.73 actions); it was also the least common behavior for learners in less engaged groups. Engaging in a full-cycle of model construction, evaluation (simulation), and revision (parameterization) led the learners to build more complex and varied models. This result aligns with findings reported in the literature such as \cite{sins2005difficult}\cite{goel2015impact}). For example, \cite{goel2015impact} describes a study in middle school science in which small teams of learners who collaboratively engaged in the evaluation and revision of the conceptual models in general created qualitatively better models than those who engaged only in the construction of conceptual models. 

Our findings also align with findings in other exploratory learning environments (for example, programming \cite{blikstein2014programming}, problem-solving \cite{hu2017learning}, video watching in MOOCS \cite{maldonado2018mining}). We were able to associate each behavior type to strategies found in other work. First, Type 1 (Observation of simulation) was associated with studying \cite{garavalia2002prior}, rehearsing \cite{broadbent2017comparing}, and only video-watching \cite{maldonado2018mining} in which learners invest time to better understand a particular concept or to attain a decent working knowledge. Second, Type 2 (Construction) is associated with testing out the connections \cite{joyner2014mila}. Type 3 (Full-cycle) is associated with problem solving behaviors of K-12 students who exhibit high meta-cognitive skills in that they constantly keep evaluating their solution by comparing their current state and the goal state \cite{hu2017learning}. Just like we focused on the behavior patterns rather than the duration of certain behaviors, Blikstein (2014) found that the changes in the code update pattern is a more important factor determining student performance in programming courses \cite{blikstein2014programming}.

\section{Conclusions}
We described a study on the use of an online laboratory for self-directed learning by constructing and simulating conceptual models of ecological systems.  In this study, we could observe only the modeling behaviors and outcomes; the learning goals and outcomes were unknown. We used machine learning techniques to analyze the modeling behaviors of 315 learners and 822 conceptual models they generated. We derive three main conclusions from the results. First, learners manifest three types of modeling behaviors: observation (simulation focused), construction (construction focused), and full exploration (model construction, evaluation and revision). Second, while observation was the most common behavior among all learners, construction without evaluation was more common for less engaged learners and full exploration occurred mostly for more engaged learners. Third, learners who explored the full cycle of model construction, evaluation and revision generated models of higher quality. These modeling behaviors provide insights into self-directed learning at large.

\subsubsection{Acknowledgements} This research was supported by US NSF grant \#1636848. We thank members of the VERA project, especially Luke Eglington and Stephen Buckley. This research was conducted in accordance with IRB protocol \#H18258. 

\bibliographystyle{splncs04}
\bibliography{mypaper}
%




\end{document}